\documentclass[twocolumn,showpacs,preprintnumbers,amsmath,amssymb]{revtex4}
\usepackage{graphicx}
\usepackage{dcolumn}
\usepackage{bm}

\begin{document}

\preprint{APS/123-QED}

\title{Shaking-induced dynamics of cold atoms in magnetic traps}

\author{I. Llorente Garc\'{i}a}
 \altaffiliation{Dept. of Physics and Astronomy, University College London, Gower St., London WC1E 6BT.}
 \email{i.llorente-garcia@ucl.ac.uk}

\author{B. Darqui\'{e}}
 \altaffiliation{Laboratoire de Physique des Lasers, Universit\'{e} Paris 13, Sorbonne Paris Cit\'{e}, CNRS, F-93430, Villetaneuse, France.}
\email{benoit.darquie@univ-paris13.fr}

\author{C. D. J. Sinclair}
 \altaffiliation{University College London Institute of Neurology, Queen Square, London WC1N 3BG, UK.}

\author{E. A. Curtis}
 \altaffiliation{National Physical Laboratory, Teddington, Middlesex, TW11 0LW, UK.}

\author{M. Tachikawa}
\altaffiliation{Dept. of Physics, Meiji University, Kawasaki-shi, Kanagawa 214-8571, Japan.}

\author{J. J. Hudson}

\author{E. A. Hinds}
\email{ed.hinds@imperial.ac.uk}

\affiliation{
Centre for Cold Matter, Imperial College London. Prince Consort Road. London SW7 2BW.
}

\date{\today}

\begin{abstract}

We describe an experiment in which cold rubidium atoms, confined in an elongated magnetic trap, are excited by transverse oscillation of the trap centre. The temperature after excitation exhibits resonance as a function of the driving frequency. We measure these resonances at several different trap frequencies. In order to interpret the experiments, we develop a simple model that incorporates both collisions between atoms and the anharmonicity of the real three-dimensional trapping potential. As well as providing a precise connection between the transverse harmonic oscillation frequency and the temperature resonance frequency,  this model gives insight into the heating and loss mechanisms, and into the dynamics of driven clouds of cold trapped atoms.

\end{abstract}

\maketitle

\section{\label{sec:Intro} Introduction}

It is common to determine the oscillation frequencies of atoms in a magnetic trap by exciting their motion. A sudden displacement of the trap excites a dipole (centre-of-mass) oscillation of the cloud at the trap frequency, while a sudden compression or de-compression induces quadrupole (width) oscillation of the cloud. In many experiments, the trapped cloud is long and thin, with transverse frequencies of 1\,kHz or above. The oscillation amplitude is then typically below the resolution of the imaging system and therefore difficult to observe directly. In that case, the oscillation may be driven, and the excitation of the cloud may be detected through the increase in its length, which results from heating. This exhibits a resonant behaviour, with a maximum temperature at a specific driving frequency, related to the natural transverse oscillation frequency.

This paper presents temperature resonances of magnetically trapped atoms, induced by dipole oscillation, with six different trap frequencies. These traps are formed on an atom chip using the field of a permanently magnetised videotape. A full characterisation of the videotape traps is carried out with particular emphasis on the effects of anharmonicity. Our experimental results are compared with a numerical model that provides valuable insight into the dynamics of the cold atoms moving in the real confining potential of the trap under the important influences of collisions and atom loss.

Several references already consider the dipolar excitation of trapped particles by shaking the trap centre \cite{st1,st2,st10,ions1,st13}. The first three are theory papers \cite{st1,st2,st10} that describe heating due to laser noise in harmonic optical traps far from resonance.  Ref. \cite{st2} calculates the evolving energy distribution, accounting for atom loss, in a truncated harmonic trap. Ref. \cite{ions1} describes measurements of the resonant frequencies and laser cooling rates for ions in a shaken Penning trap. In Ref.\,\cite{st13} M. Kumakura et al.  investigate the excitation of neutral atoms in a shaken cloverleaf magnetic trap. They measure resonances in atom loss and temperature and discuss how different ratios of atomic temperature to effective trap depth result in either heating or cooling after shaking. They use a classical 1D equation of motion, without collisions, to illustrate some features of their experiment.

There have also been discussions of parametric excitation, i.e. modulation of the trap frequency, in various contexts. These include heating and cooling of neutral atoms in optical dipole traps \cite{st1,st2,st10,st3,st6,st5}, one-dimensional optical lattices \cite{st4,st6,st10} or magnetic traps \cite{st17}, and measurement of oscillation frequencies for electrons in a Penning trap \cite{st11,st12}, ions in a quadrupole trap \cite{st16} or neutral atoms in a MOT \cite{st14,st15}. We do not study parametric resonance here because there is no straightforward way to modulate the frequency of our videotape trap without also shaking its position. The same is true of any permanent-magnet atom trap.

The experiments described here determine both the position and the shape of the dipolar temperature resonances. We also investigate these resonances theoretically with the aid of a three-dimensional numerical model that accounts simultaneously for anharmonicity of the trap, atom loss due to finite trap depth, interatomic collisions and evolution of the collision rate during the excitation. All these aspects of our model go beyond what has been done before. Our model yields good quantitative agreement between experiment and theory, even though we use a rather simplified collision model. We anticipate that our simplifying assumption could be exported to a number of ensemble dynamic problems in order to achieve faster simulations.

\section{\label{sec:experim} The experiment}

\begin{figure*}
\begin{minipage}{0.3\linewidth}
\centering
\includegraphics[width=0.95\textwidth]{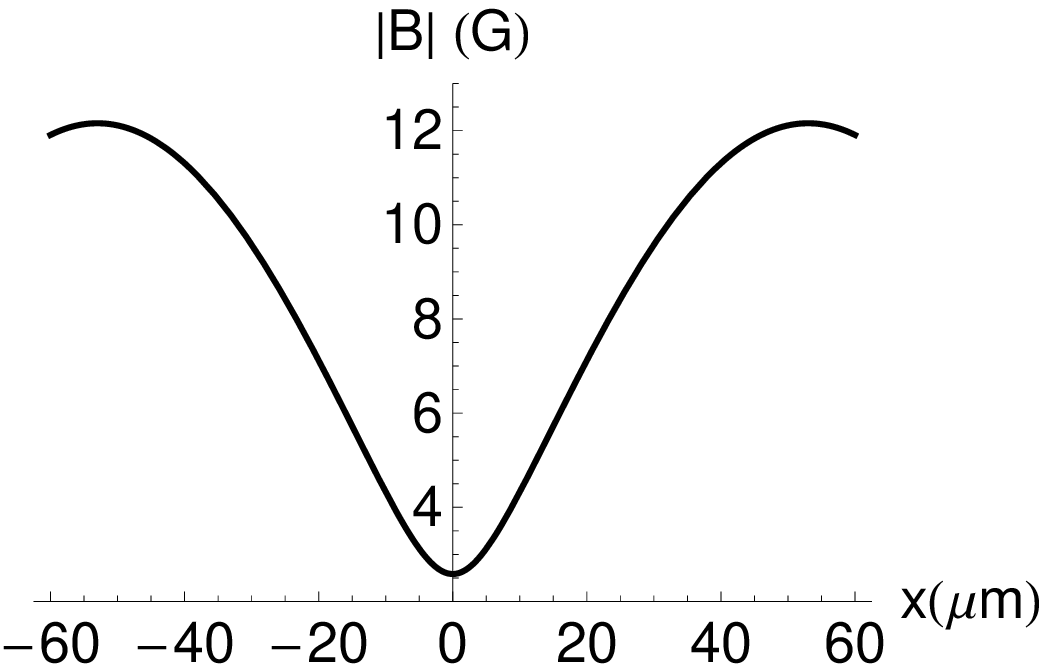}
\end{minipage}
\hspace{0.1cm}
\begin{minipage}{0.3\linewidth}
\centering
\includegraphics[width=0.95\textwidth]{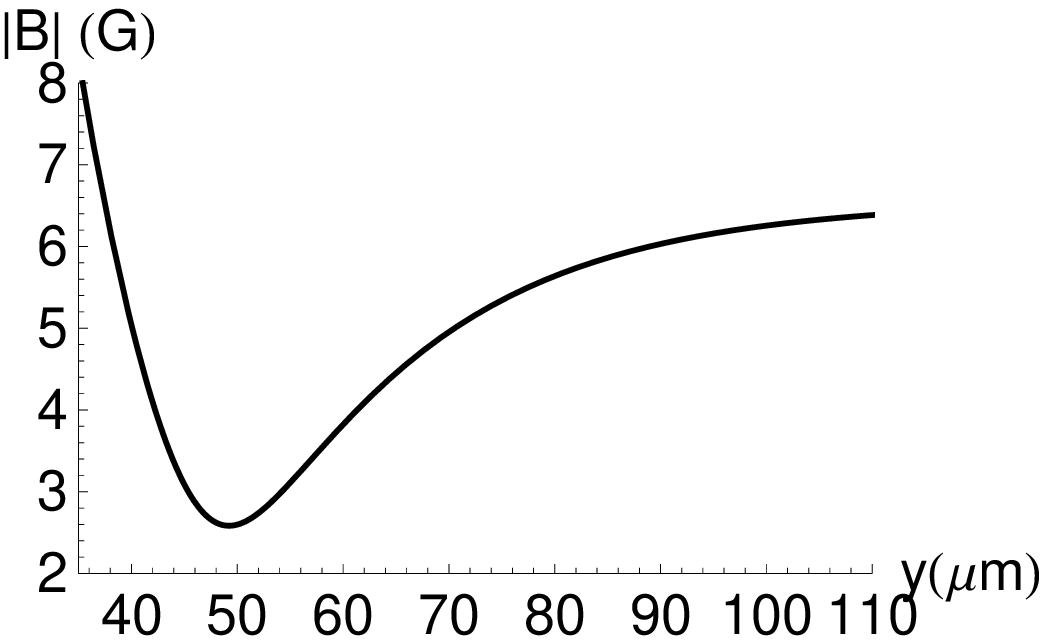}
\end{minipage}
\hspace{0.1cm}
\begin{minipage}{0.3\linewidth}
\centering
\includegraphics[width=0.95\textwidth]{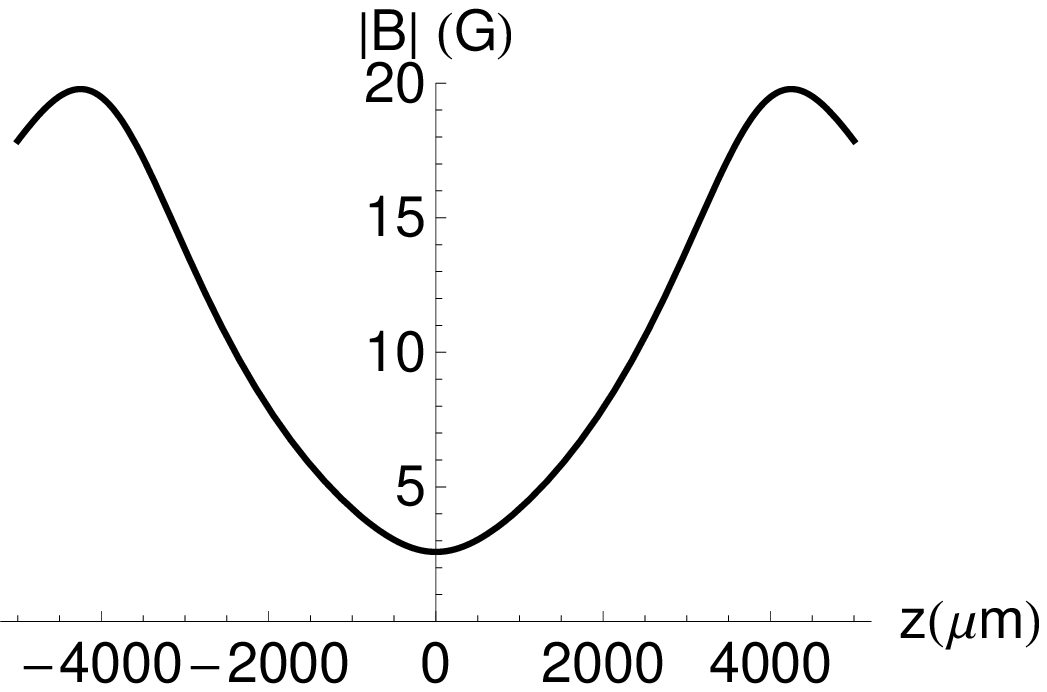}
\end{minipage}
\\
\vspace{0.3cm}
\begin{minipage}{0.3\linewidth}
\centering
\includegraphics[width=0.95\textwidth]{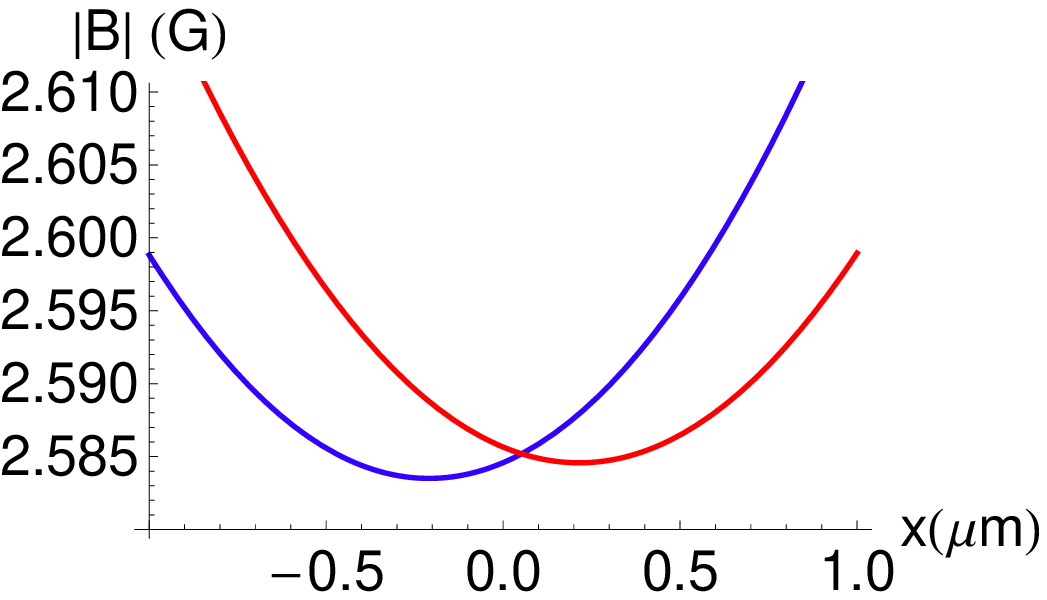}
\end{minipage}
\hspace{0.1cm}
\begin{minipage}{0.3\linewidth}
\centering
\includegraphics[width=0.95\textwidth]{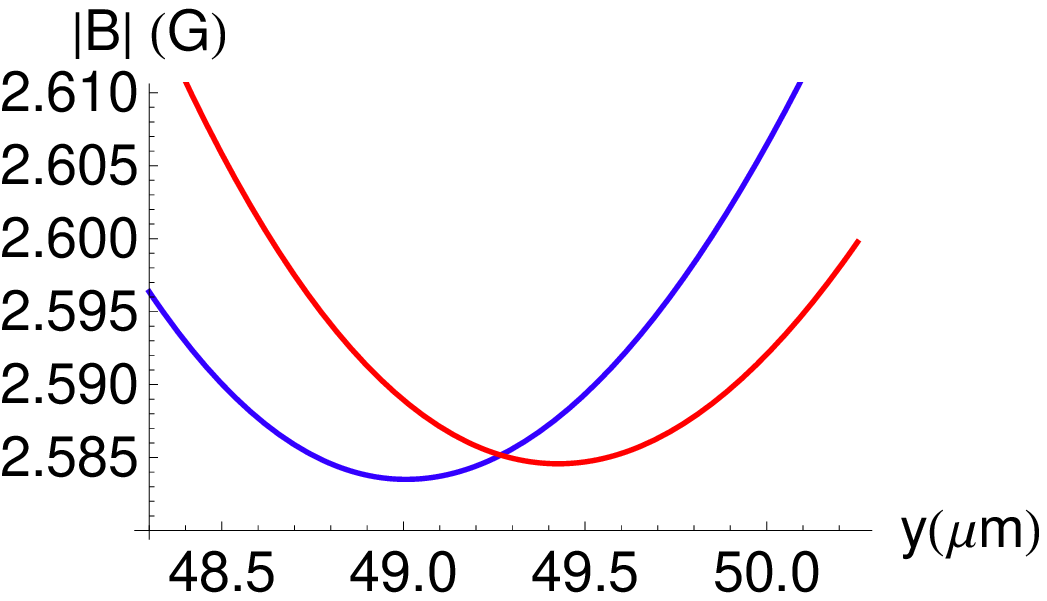}
\end{minipage}
\hspace{0.1cm}
\begin{minipage}{0.3\linewidth}
\centering
\includegraphics[width=0.95\textwidth]{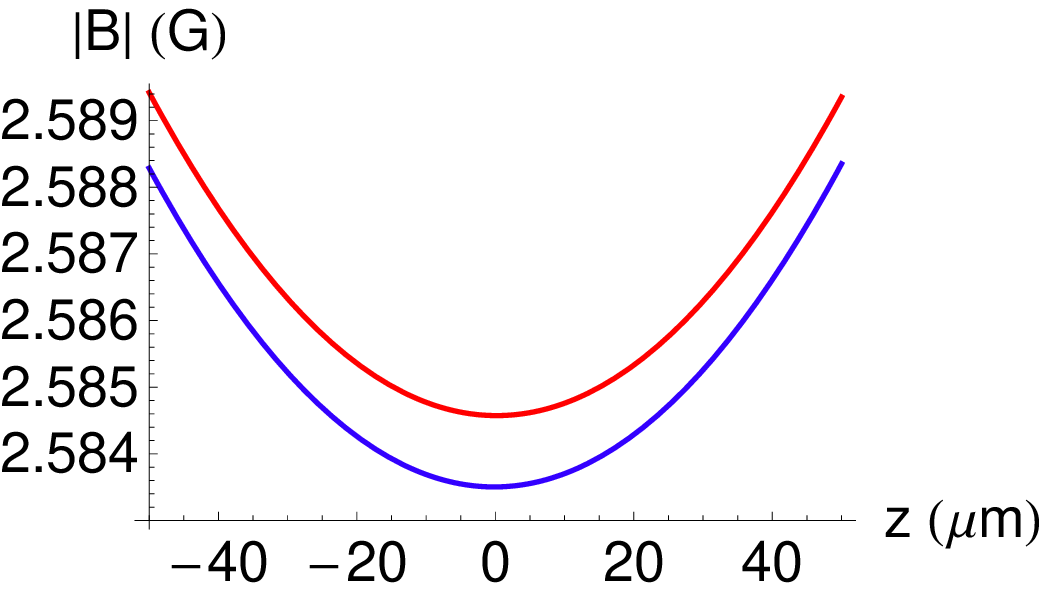}
\end{minipage}
\caption{Total field strength of magnetic trap versus position. The transverse bias field is $B_b=5.9\,\mathrm{G}$ along $\hat{x}$. Upper row: overview along each Cartesian direction. Lower row: zoom into the centre shows the extrema of the modulation induced by the shaking field $b=105\,$mG.}
\label{fig:howPotChanges}
\end{figure*}

Our experiment uses the field gradient of a magnetized videotape to trap the atoms. The tape, lying in the $xz$ plane, has magnetisation $M_1\cos(kx)\hat{x}$, where $k=(2\pi/110)\,\mu\mbox{m}^{-1}$. Above the surface, this magnetisation  produces a transverse (i.e. in the $xy$ plane) magnetic field of $\{B_{x},B_{y}\}=B_{1}e^{-k y}\{-\cos(kx),\sin(ky)\}$ \cite{HindsHughesReview} with $B_1\simeq 100\,$G.  The addition of a transverse bias field, $B_{b}\{\cos(\theta),\sin(\theta)\}$, produces an array of zeros in the total transverse field located at a height $y=\ln(B_{1}/B_{b})/k$. On Taylor expanding the field around any of these zeros one finds a quadrupole field in the $xy$ plane whose orientation depends on the angle $\theta$ of the bias field. The magnitude of the field grows linearly with the cylindrical radius $r$ according to $|B|=k B_{b} r$, providing a cylindrically symmetrical, linear confining potential $\mu_{B}g_{F}m_{F}|B|$ for weak-field seeking atoms with magnetic quantum number $m_{F}$ and $g$-factor $g_{F}$ ($\mu_B$ is the Bohr magneton).  With the addition of a further bias field $B_{z}\hat{z}$, the trapping potential for transverse displacements becomes approximately harmonic in the region close to the axis, where $k B_{b} r\ll B_{z}$. In that region, atoms oscillate in the trap with a transverse frequency of
\begin{equation}\label{eqn:radialVideofreq}
f_{r}\approx \frac{k B_b}{2\pi}\sqrt{\frac{\mu_B g_F m_F}{m
B_{z}}}\, ,
\end{equation}
where $m$ is the mass of the atom. We use $^{87}$Rb atoms in the $(F=2, m_{F}=2)$ ground state, for which $g_{F}m_{F}=1$. The bias field $B_{b}$ is tuneable over the range $6-36$\,G, which varies the distance from the trap to the videotape between $50$ and $18\,\mu$m. Over this range of positions the axial bias $B_{z}$ drops from $2.58$ to $2.51\,$G. These parameters give transverse harmonic oscillation frequencies in the range $3-20\, \mathrm{kHz}$. The axial bias field has a minimum value near the centre of the videotape and increases roughly quadratically with $z$ to form a weak axial trap with a frequency of $15\,$Hz. In summary, this combination of videotape fields and bias fields produces an array of cigar-shaped 3D traps with strong transverse confinement and weaker trapping along $\hat{z}$. For more details of videotape traps, see \cite{HindsHughesReview,myVideoPaper}.

The upper graphs in Figure\,\ref{fig:howPotChanges} show how the total magnetic field strength $|B|$ varies along the three Cartesian axes through the centre of the trap. These are calculated using a full numerical model of the apparatus, with $B_b=5.9\,$G, directed along $\hat{x}$. The trap along $x$ repeats every $110\,\mu$m because of the periodic magnetisation of the videotape, but only one of these is shown as only one trap is used in the experiments. Along $y$, we see an asymmetric trap, with a strong repulsive wall as the videotape is approached and an asymptote far from the tape that is equal to the total bias field strength. Along $z$, there is the weak trapping due to the inhomogeneous axial bias field.

We load one such trap using an experimental sequence similar to that described in \cite{myVideoPaper}. The temperature of the atom cloud depends on the bias field and ranges from $13\,\mu$K at $B_b=5.9\,$G up to $130\,\mu$K at $B_b=35.6\,$G. We then add a modulation field $b \cos(2\pi f t)$ along $\hat{x}+\hat{y}$, to displace the trap by a small distance $-\tfrac{b}{k B_b}\cos(\omega t)$ along $\hat{x}+\hat{y}$. We choose $b=105\,$mG, which gives a shaking amplitude of $50-300\,$nm over the range of transverse bias fields used. Zooming into the centre of the trap, the three lower graphs in Fig.\,\ref{fig:howPotChanges} show the extrema of this modulation. The first two show the (equal) movements along $x$ and $y$ resulting from our particular arrangement of fields. The third shows the absence of movement along $z$ and also illustrates the modulation of the minimum field, due to the variation of $B_z$ with transverse displacement of the trap. Since $B_z$ affects the transverse trap frequency, this could cause parametric heating, but the effect is considered in \cite{MyThesis} and found to be negligible.

\begin{figure}[b]
\includegraphics[width=0.4\textwidth]{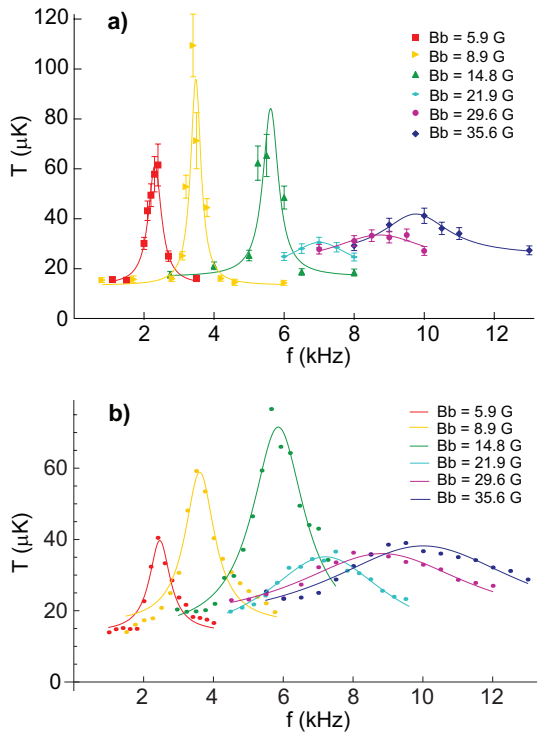}
\caption{Temperature of cloud after shaking for $5\,$s, plotted as a function of shaking frequency $f$. (a) Experimental results. The points are measurements taken at six transverse bias fields, $B_b$. Lines are Lorentzian fits to the data.  (b) Simulation. Points are calculated and lines are Lorentzian fits.}
\label{fig:resonances}
\end{figure}

With $B_b$ initially set at $5.9\,$G, the atoms are shaken for $5\,\mathrm{s}$ and held for $1\,$s, after which the temperature of the trapped cloud is determined by measuring its density distribution along the $z$-direction. This is done using a CCD camera to record the absorption of resonant laser light \cite{myVideoPaper}, taking into account the inhomogeneous Zeeman shift of the trapped atoms. The experiment is repeated for a range of shaking frequencies to map out a resonance curve, shown in the leftmost peak of Fig. \ref{fig:resonances}(a). The next two curves are obtained in the same way with $B_b=8.9\,$G and $B_b=14.8\,$G. At still higher transverse bias fields ($B_b=21.9, 29.6$ and $35.6\,$G), the atoms are too close to the videotape to yield clean absorption images. In these three cases the trap is moved away from the surface by lowering the bias to $3\,$G over $3\,$s before taking the image. This weakens the trap, thereby cooling the cloud, and it is the lower temperature in this final trap that we plot in Fig. \ref{fig:resonances}(a).

\begin{figure}[b]
  \includegraphics[width=0.9\columnwidth]{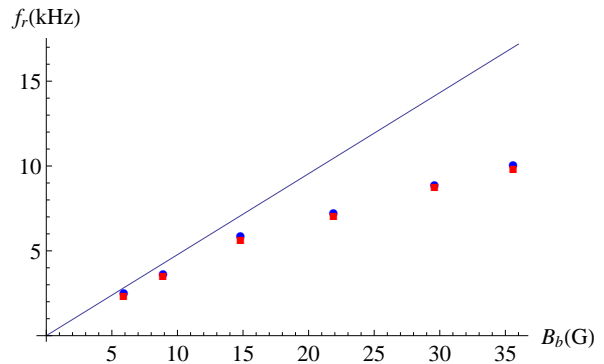}
  \caption{Frequency of temperature resonance versus bias magnetic field. Solid line: calculated harmonic frequency of transverse oscillations at trap centre. Red squares: measured centre frequencies of the temperature resonances in Fig.\,\ref{fig:resonances}(a). Blue circles: centre frequencies of the simulated temperature resonances plotted in Fig.\,\ref{fig:resonances}(b).}
  \label{fig:FreqVsBias}
\end{figure}

These temperature resonances show that the atoms absorb energy most efficiently near a particular frequency. In the case of a one-dimensional harmonic trap, the physics would be that of a driven, weakly damped harmonic oscillator, whose Lorentzian resonance would be centred on the oscillator frequency with a width given by the collision rate. The actual widths are much greater than the collision rate, but nevertheless, motivated by this thought, we fit a Lorentzian to each resonance curve and plot the centre frequencies (the red squares) as a function of $B_b$ in Fig.\,\ref{fig:FreqVsBias}. For comparison, the line in Fig.\,\ref{fig:FreqVsBias} shows the frequency of small transverse oscillations, given by Eq.\,\ref{eqn:radialVideofreq}, with $B_z$ evaluated at the centre of the trap for each value of transverse bias.  The data points lie below this line because (i) the most energetic atoms move out of the region of small ${x,y}$ where the harmonic approximation of Eq.\,(\ref{eqn:radialVideofreq}) is valid and (ii) the atoms are also displaced from the centre along $z$, where the increased value of $B_z$ reduces the radial frequency. These effects also produce inhomogeneous broadening of the temperature resonances, making them wider for the hotter clouds at higher bias fields, as seen in Fig.\,\ref{fig:resonances}(a).

\begin{figure}[t]
\begin{minipage}{0.9\linewidth}
\centering
\includegraphics[width=0.75\textwidth]{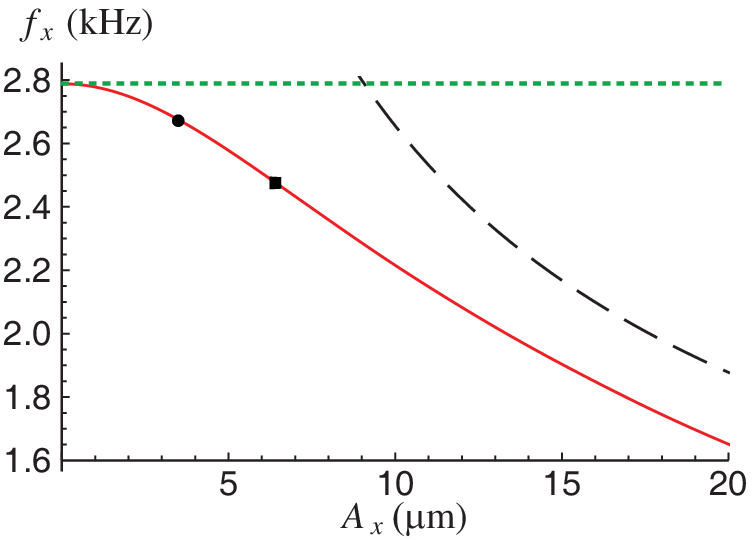}
\end{minipage}
\hspace{0.5cm}
\begin{minipage}{0.9\linewidth}
\centering
\includegraphics[width=0.75\textwidth]{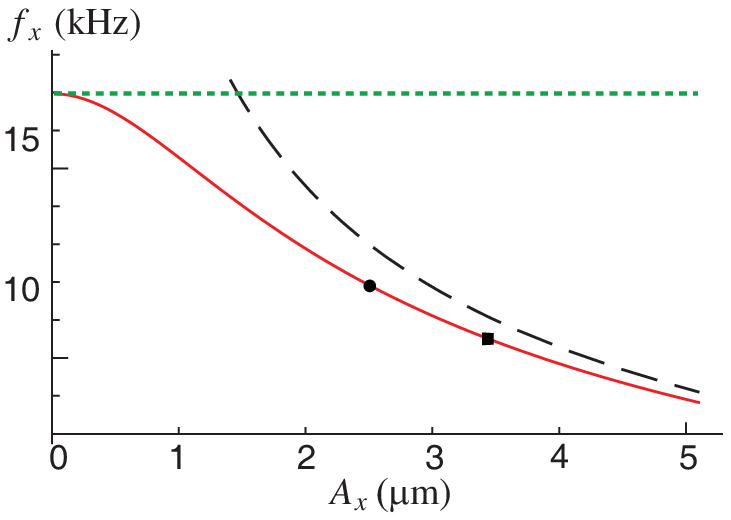}
\end{minipage}
\caption{Frequency of oscillations along $\hat{x}$ through centre of trap as a function of amplitude. Upper: $B_b=5.9\,\mathrm{G}$. Lower: $B_b=35.6\,\mathrm{G}$. Solid red line: frequency calculated from numerical solution to full equation of motion. Dotted green line: Harmonic approximation. Dashed black line: linear potential approximation. Circles: rms cloud radius at initial temperature.  Squares: rms cloud radius after heating at resonance.}
\label{fig:FreqsVsAmpliBias2and7}
\end{figure}

\begin{figure}[t]
  \includegraphics[width=0.36\textwidth]{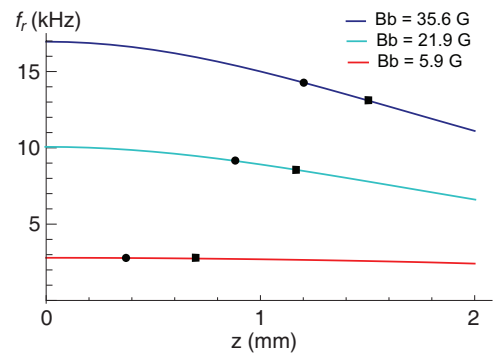}
  \caption{Frequency of small (harmonic) transverse oscillations as a function of displacement $z$ from the trap centre. \textbf{Circles}: rms cloud radius (half-length) at initial temperature.  \textbf{Squares}: rms cloud radius (half-length) after heating at resonance.}
  \label{fig:RadialFreqVsZ}
\end{figure}

Numerically integrating the equations of motion in the full trap potential, we have calculated the period of oscillation along the $x$ direction through the centre of the trap. The inverse of this is plotted as a function of amplitude $A_x$  by the solid lines in Fig.\,\ref{fig:FreqsVsAmpliBias2and7}, the upper(lower) panel being for the case of $B_b=5.9(35.6)\,$G. At small amplitude, the frequency coincides with the harmonic approximation indicated by the dotted line. At large amplitude, the potential approaches that of  linear trap, and the frequency tends correspondingly to $\tfrac{1}{4}\sqrt{\mu_B k B_b/(2 m A_x)}$, indicated by the dashed line. The filled circles(squares) mark the rms radius of the cloud in this direction before(after) resonant heating. These show that the atoms explore the anharmonic region of the transverse trap even before the cloud is heated, and move further into this region after heating. These frequency shifts are larger when the bias field is larger. Figure\,\ref{fig:RadialFreqVsZ} shows how \emph{axial} displacement from the centre of the trap lowers the harmonic frequency for small transverse oscillations. The essence of this effect is already captured in Eq.\,(\ref{eqn:radialVideofreq}) through the dependence of $f_r$ on $B_z$, but here we show the result of the full numerical model of our experiment. Again, the circles(squares) represent the rms size, this time along $z$ before(after) heating. Both of these mechanisms contribute appreciably to the inhomogeneous broadening seen in Fig.\,\ref{fig:resonances}(a) and the lowering of the trap frequency seen in Fig.\,\ref{fig:FreqVsBias}.

For a more quantitative understanding of the resonances it is necessary to build a dynamical model that allows the atoms to collide. In the next section we develop such a model and use it to simulate the data presented in Fig.\,\ref{fig:resonances}(a).

\section{\label{sec:simul} Simulation}
\subsection{\label{sec:simul:a} The role of collisions}
Our goal here is to find a numerical model that is simplified as far as possible, while still reproducing the data of Fig.\,\ref{fig:resonances}(a). We use the known currents and videotape magnetisation to determine an accurate 3D  potential,  $U(x,y,z,t)$, for the shaken magnetic trap. We note that the inclusion of gravity has no significant effect because these traps are so strong vertically. The cloud of approximately $10^5$ atoms is represented by an ensemble of 500-5000 point particles, this being sufficient to represent the average properties of the ensemble. The three initial velocity components for each particle are chosen at random from the Maxwell-Boltzmann distribution corresponding to initial temperature $T_i$, which is the baseline temperature in Fig.\,\ref{fig:resonances}(a) for that particular trap. Similarly, the initial positions are distributed with a probability density proportional to $\exp [-U(x,y,z,0)/(k_{B}T_{i})]$, where $k_{B}$ is the Boltzmann constant. The classical equations of motion are then integrated numerically to follow the movement of the particles.

\begin{figure}[t]
\begin{minipage}{0.65\linewidth}
\centering
\includegraphics[width=0.99\textwidth]{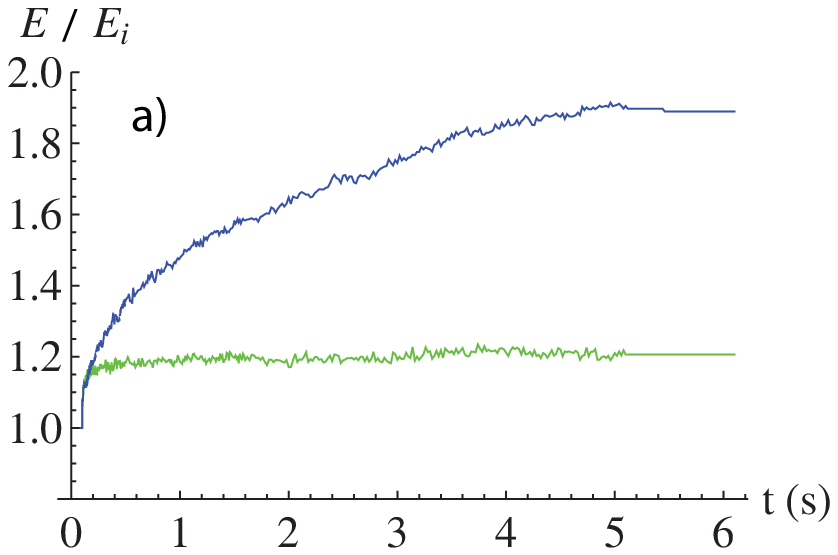}
\end{minipage}\\
\begin{minipage}{0.65\linewidth}
\centering
\includegraphics[width=0.99\textwidth]{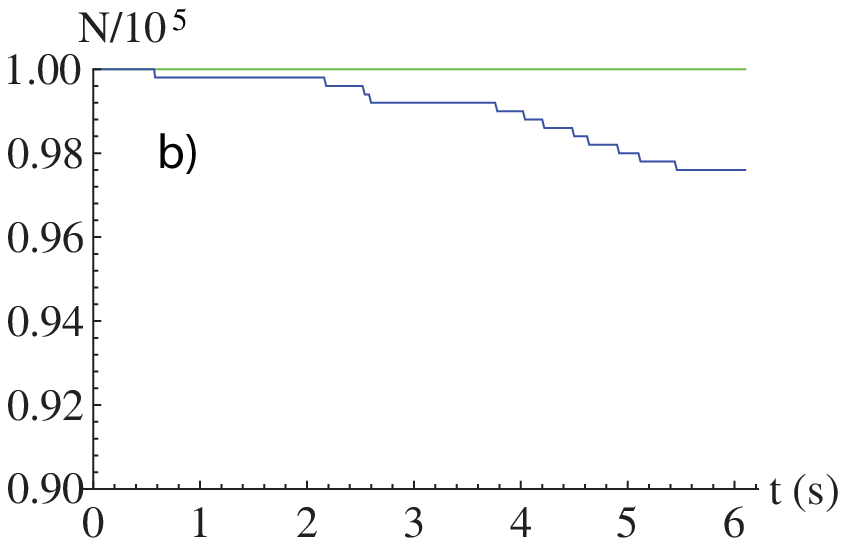}
\end{minipage}\\
\caption{Simulation of $10^5$ atoms in a trap having $B_b=21.9\,\mathrm{G}$.  The trap is shaken by a modulation field of  $105\,$mG amplitude and  $7.2\,\mathrm{kHz}$ frequency over the time interval  $t=0.1-5.1\,$s. The initial temperature is $71\,\mu \mathrm{K}$. These parameters correspond to the resonant peak of the pale blue curve in Fig.\,\ref{fig:resonances}(a). This simulation uses 500 particles. Green curves: no collisions. Blue curves: momentum is redistributed through collisions. (a) Increase of total energy with time. (b) Number of atoms remaining within the volume of the trap, with the initial atom number being $N_i=10^5$.}
\label{fig:peak5}
\end{figure}

The two curves in Fig.\,\ref{fig:peak5}(a) show how the energy of the cloud increases with time when the trap having $B_b = 21.9\,$G  is shaken at 7.2kHz. These parameters correspond to the peak of the pale blue resonance curve at $7\,$kHz in Fig.\,\ref{fig:resonances}(a). A simple model without collisions produces the green (lower) curve in Fig.\,\ref{fig:peak5}(a). Particles are placed in the trap at time $t=0$, which remains static for the first $0.1\,$s. Then the shaking is switched on and the energy rises rapidly, increasing by $20\%$ over the next $0.1\,$s. This is  due to the excitation of particles whose transverse oscillation frequency is close to the drive frequency.  Once they are sufficiently excited, the anharmonicity moves these particles out of resonance. Because the resonant group has been depleted, there is almost no subsequent energy increase even though the shaking continues until $t=5.1\,$s. The shaking is then switched off leaving the cloud to evolve freely over the last second of the simulation. This behaviour disagrees with the experiment. In reality, the temperature doubles (though not in Fig.\,\ref{fig:resonances}(a) because there the trap was relaxed before recording the temperature). The discrepancy is removed when we allow the simulation to redistribute the momentum through collisions.

The following very simple model of momentum redistribution is sufficient for our purpose. The thermally averaged atom-atom scattering cross-section is $\sigma=8 \pi a^{2}/(1+2\pi a^2 m k_{B} T /\hbar^2)$, where $a=5.53\times10^{-9}\,\mathrm{m}$ is the s-wave scattering length \cite{scatteringLength} and $T$ is the temperature of the cloud. The average relative velocity between particles is $v = \sqrt{16 k_{B} T/(\pi m)}$ \cite{PhysChem}. Knowing the mean density of trapped atoms, $ n $, we obtain a mean collision rate per atom,  $ n \sigma v$. At appropriate time intervals (short compared with the inverse collision rate), the numerical integration is paused and a random number is generated for each particle to determine whether it has had a collision. Ideally, this should account for the local variations of density \cite{Bird94}, which we ignore here in order to keep the model simple. If there is no collision, the atom continues unperturbed. Otherwise, the momentum of the particle is redirected by the collision, according to some angular distribution. For the data presented here, we used that of elastic scattering from an infinitely heavy sphere. However, we find that the results are quite insensitive to the chosen distribution and there is nothing special about this particular one. Given our experimental conditions, the average time between collisions is $30-100\, \mathrm{ms}$ at the start of the excitation, and this becomes longer as the cloud heats up.

With the momentum redistribution thus incorporated,  we obtain the dark blue (higher) curve in Fig.\,\ref{fig:peak5}(a).  Now, the ensemble is able to continue absorbing energy after the initial absorption because the depleted velocity group is steadily replenished through collisions. The refilling rate slows down as the atoms become more energetic, and this is responsible for the saturation of heating, seen in Fig.\,\ref{fig:peak5}(a). In this case, the energy doubles over the $5\,$s of shaking and our model approximates well  the measured heating of the atom cloud.

Figure \,\ref{fig:peak5}(b) plots the number of atoms in the trap as a function of time, with collisions (blue, lower) and without (green, higher). In the collision-free case, few atoms leave the trap because the heating is weak and because the phase space is not efficiently sampled in the absence of collisions.  By contrast, when collisions are included, the cloud heats much more strongly and the energetic atoms are more easily able to find an exit route from the trap. There is competition between the heating rate due to resonant excitation and the cooling rate due to evaporation from the trap. In this example, the heating is dominant because the trap is deep in comparison with the mean energy absorbed by each atom.

\begin{figure}
 \includegraphics[width=0.4\textwidth]{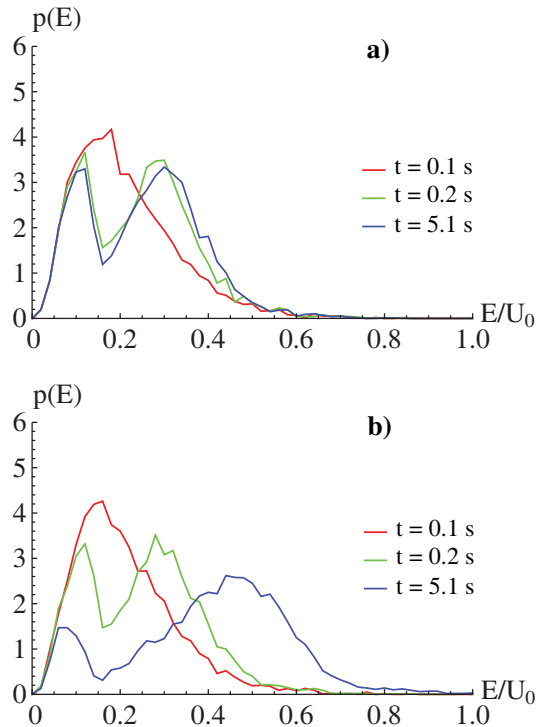}
 \caption{Simulated energy spectra of $10^5$ atoms, initially at $71\,\mu \mathrm{K}$ in a trap having $B_b=21.9\,\mathrm{G}$. They are shaken over the time interval  $t=0.1-5.1\,$s by a $7.2\,\mathrm{kHz}$  field of  amplitude $b=105\,$mG. These are the same parameters used in Fig.\,\ref{fig:peak5} and at the resonant peak of the pale blue curve in Fig.\,\ref{fig:resonances}(a). This simulation uses 5000 particles. (a) Without collisions. (b) With collisions. Red curves: Initial thermal distribution just before shaking. Green curves: After $0.1\,$s of shaking. Blue curves: After $5\,$s of shaking.}
 \label{fig:energySpectra}
\end{figure}

Figure \ref{fig:energySpectra}(a) shows calculated energy spectra in the absence of collisions. The ordinate is the probability density for a given energy, normalised to unity, while the abscissa shows that energy,  normalised to the trap depth $U_0=1.2\,$mK. The red curve ($t=0.1\,$s), showing the initial distribution just before the trap starts to shake, has a single peak just below $E=0.2 U_0$. After only $100\,$ms of shaking (green curve, $t=0.2\,$s) a deep notch appears in the distribution close to the energy of the initial peak. This shows that atoms close to that energy, having an oscillation period close to the period of the drive, are the ones excited by the shaking. Their excitation causes a second peak in the distribution at approximately $E/U_{0}=0.3$. There is no further significant change in the distribution, even after $5\,$s of shaking, as shown by the blue curve. This behaviour is to be compared with Figure \ref{fig:energySpectra}(b), which shows energy distributions for the same simulated experiment when collisions are included. The initial distribution is the same, as is the notch appearing at $0.2\,$s, but in this case continued shaking does produce additional heating because the collisions refill the velocity group that absorbs energy from the drive. This is clearly seen in the growing probability on the high-energy end of the spectrum.

\begin{figure}[t]
  \includegraphics[width=0.4\textwidth]{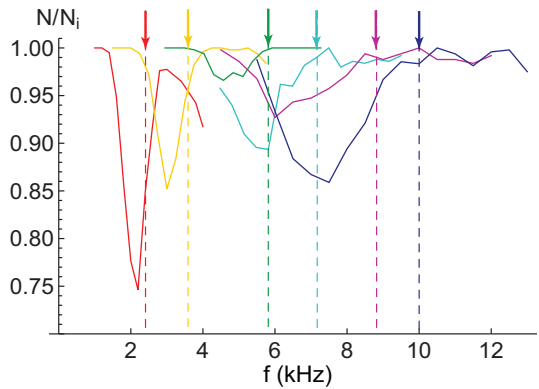}\\
  \caption{Simulated ratio of final atom number $N$ to initial number $N_i$ as a function of the excitation frequency. The colour code corresponds to that of Fig. \ref{fig:resonances}, with coloured arrows marking the temperature resonance frequencies. For the lowest three bias fields, the lowest barrier to escape is along \emph{y} ($360\, \mu \mathrm{K}$, $520\, \mu \mathrm{K}$ and $880\, \mu \mathrm{K}$ in order of increasing bias field), while for the highest three, it is along \emph{z} ($\sim 1200\, \mu \mathrm{K}$). For comparison, the measured initial cloud temperatures before the excitation take values of $13.4\, \mu \mathrm{K}$, $13.3\, \mu \mathrm{K}$, $16.7\, \mu \mathrm{K}$, $71\, \mu \mathrm{K}$, $98\, \mu \mathrm{K}$ and $132\, \mu \mathrm{K}$, in order of increasing bias field.}
  \label{fig:simulatedResonAtomLoss}
\end{figure}

\subsection{\label{sec:simul:b} Simulated Resonances}

Figure \ref{fig:resonances}(b) shows our simulated temperature resonances at each of the six different values of the bias field. After the shaking stops in the experiment, we allow the atoms to thermalise for one second before measuring the temperature. In the simulation, we simply take the final energy as a measure of the temperature that would be reached in equilibrium. The central frequencies of these simulated resonances agree very closely with the experiment, as indicated by the blue circles in Fig.\,\ref{fig:FreqVsBias}.

The peak temperature rises are also remarkably well reproduced by the simulations, given the simplicity of the model. In particular, these model collisions redistribute momentum, but do not permit re-thermalisation of the energy distribution because the energy of a given atom is conserved in the collision. Despite that, the energy increase in the model reproduces all the measured temperature rises to well within a factor of two and that remains the case for a variety of model angular distributions.

The widths of the simulated resonant peaks are a little too large - by a factor of $1.5 - 2$. In searching for an explanation we simulated the yellow ($8.9\,$G) resonance with the initial temperature reduced from $13.3\,\mu$K to $7\,\mu$K. This only reduced the width by $10\%$, so we do not think a temperature calibration error can explain the discrepancy between the measured and calculated widths. It could well be that our very simple collision model causes the resonances to be too broad, although we do not see a clear reason why that should be so.

Figure \ref{fig:simulatedResonAtomLoss} shows the simulated resonances in atom loss for the same six bias fields used in Fig.\,\ref{fig:resonances}. We see that the dips in atom number are on the low-frequency side of the temperature resonances indicated by arrows. This is because the energetic atoms most likely to be driven out of the trap are also those most able to explore the anharmonic regions and hence to oscillate at  lower frequencies. The same behaviour has been reported by several groups for both parametric shaking \cite{st4,st5,st9,st17,st19}, and for shaking of the trap centre \cite{st13}, in agreement with the results of our simulations.\\

\section{\label{sec:Conclusions} Summary and Conclusions}
We have measured the temperature rise in a cigar-shaped cloud of cold atoms after shaking it sinusoidally in the transverse direction. Unlike most previous measurements, we have modulated the position of the trap, not its curvature. We have recorded temperature resonances as a function of frequency for several values of the bias field that controls the curvature of the transverse trapping potential.  Essential to the interpretation of our measurements is the understanding that,  at the temperatures involved, atoms oscillate with a wide range of transverse frequencies in the trap and hence that the central frequencies of the observed resonances lie below the calculated harmonic frequencies for small oscillations.

We have developed a simple numerical model that has provided a clear quantitative understanding of the driven dynamics. When compared with other models and simulations in the literature, this provides one of the most detailed and complete attempts to reproduce the observed resonances. No other models include both the collisions between particles and the anharmonicity of the trapping potential, yet both of these are shown to be essential for a full understanding of the behaviour.

\begin{acknowledgments}
The authors thank Robert Nyman and Michael Trupke for useful discussions. We are indebted to the FastNet and AtomChips European networks and to the UK EPSRC and Royal Society for their funding support.
\end{acknowledgments}

%
%

\bibliography{BiblioDatabase}

\begin{thebibliography}{23}
\expandafter\ifx\csname natexlab\endcsname\relax\def\natexlab#1{#1}\fi
\expandafter\ifx\csname bibnamefont\endcsname\relax
  \def\bibnamefont#1{#1}\fi
\expandafter\ifx\csname bibfnamefont\endcsname\relax
  \def\bibfnamefont#1{#1}\fi
\expandafter\ifx\csname citenamefont\endcsname\relax
  \def\citenamefont#1{#1}\fi
\expandafter\ifx\csname url\endcsname\relax
  \def\url#1{\texttt{#1}}\fi
\expandafter\ifx\csname urlprefix\endcsname\relax\def\urlprefix{URL }\fi
\providecommand{\bibinfo}[2]{#2}
\providecommand{\eprint}[2][]{\url{#2}}

\bibitem[{\citenamefont{Savard et~al.}(1997)\citenamefont{Savard, O'Hara, and
  Thomas}}]{st1}
\bibinfo{author}{\bibfnamefont{T.~A.} \bibnamefont{Savard}},
  \bibinfo{author}{\bibfnamefont{K.~M.} \bibnamefont{O'Hara}},
  \bibnamefont{and} \bibinfo{author}{\bibfnamefont{J.~E.}
  \bibnamefont{Thomas}}, \bibinfo{journal}{Phys. Rev. A}
  \textbf{\bibinfo{volume}{56}}, \bibinfo{pages}{R1095} (\bibinfo{year}{1997}).

\bibitem[{\citenamefont{Gehm et~al.}(1998)\citenamefont{Gehm, O'Hara, Savard,
  and Thomas}}]{st2}
\bibinfo{author}{\bibfnamefont{M.~E.} \bibnamefont{Gehm}},
  \bibinfo{author}{\bibfnamefont{K.~M.} \bibnamefont{O'Hara}},
  \bibinfo{author}{\bibfnamefont{T.~A.} \bibnamefont{Savard}},
  \bibnamefont{and} \bibinfo{author}{\bibfnamefont{J.~E.}
  \bibnamefont{Thomas}}, \bibinfo{journal}{Phys. Rev. A}
  \textbf{\bibinfo{volume}{58}}, \bibinfo{pages}{3914} (\bibinfo{year}{1998}).

\bibitem[{\citenamefont{J\'{a}uregui}(2001)}]{st10}
\bibinfo{author}{\bibfnamefont{R.}~\bibnamefont{J\'{a}uregui}},
  \bibinfo{journal}{Phys. Rev. A} \textbf{\bibinfo{volume}{64}},
  \bibinfo{pages}{053408} (\bibinfo{year}{2001}).

\bibitem[{\citenamefont{van Eijkelenborg et~al.}(1999)\citenamefont{van
  Eijkelenborg, Dholakia, Storkey, Segal, and Thompson}}]{ions1}
\bibinfo{author}{\bibfnamefont{M.~A.} \bibnamefont{van Eijkelenborg}},
  \bibinfo{author}{\bibfnamefont{K.}~\bibnamefont{Dholakia}},
  \bibinfo{author}{\bibfnamefont{M.~E.~M.} \bibnamefont{Storkey}},
  \bibinfo{author}{\bibfnamefont{D.~M.} \bibnamefont{Segal}}, \bibnamefont{and}
  \bibinfo{author}{\bibfnamefont{R.~C.} \bibnamefont{Thompson}},
  \bibinfo{journal}{Optics Communications} \textbf{\bibinfo{volume}{159}},
  \bibinfo{pages}{169} (\bibinfo{year}{1999}).

\bibitem[{\citenamefont{Kumakura et~al.}(2003)\citenamefont{Kumakura,
  Shirahata, Takasu, Takahashi, and Yabuzaki}}]{st13}
\bibinfo{author}{\bibfnamefont{M.}~\bibnamefont{Kumakura}},
  \bibinfo{author}{\bibfnamefont{Y.}~\bibnamefont{Shirahata}},
  \bibinfo{author}{\bibfnamefont{Y.}~\bibnamefont{Takasu}},
  \bibinfo{author}{\bibfnamefont{Y.}~\bibnamefont{Takahashi}},
  \bibnamefont{and} \bibinfo{author}{\bibfnamefont{T.}~\bibnamefont{Yabuzaki}},
  \bibinfo{journal}{Phys. Rev. A} \textbf{\bibinfo{volume}{68}},
  \bibinfo{pages}{021401(R)} (\bibinfo{year}{2003}).

\bibitem[{\citenamefont{Gardiner et~al.}(2000)\citenamefont{Gardiner, Ye,
  Nagerl, and Kimble}}]{st3}
\bibinfo{author}{\bibfnamefont{C.~W.} \bibnamefont{Gardiner}},
  \bibinfo{author}{\bibfnamefont{J.}~\bibnamefont{Ye}},
  \bibinfo{author}{\bibfnamefont{H.~C.} \bibnamefont{Nagerl}},
  \bibnamefont{and} \bibinfo{author}{\bibfnamefont{H.~J.}
  \bibnamefont{Kimble}}, \bibinfo{journal}{Phys. Rev. A}
  \textbf{\bibinfo{volume}{61}}, \bibinfo{pages}{045801}
  (\bibinfo{year}{2000}).

\bibitem[{\citenamefont{Friebel et~al.}(1998)\citenamefont{Friebel, D'Andrea,
  Walz, Weitz, and H\"{a}nsch}}]{st6}
\bibinfo{author}{\bibfnamefont{S.}~\bibnamefont{Friebel}},
  \bibinfo{author}{\bibfnamefont{C.}~\bibnamefont{D'Andrea}},
  \bibinfo{author}{\bibfnamefont{J.}~\bibnamefont{Walz}},
  \bibinfo{author}{\bibfnamefont{M.}~\bibnamefont{Weitz}}, \bibnamefont{and}
  \bibinfo{author}{\bibfnamefont{T.~W.} \bibnamefont{H\"{a}nsch}},
  \bibinfo{journal}{Phys. Rev. A} \textbf{\bibinfo{volume}{57}},
  \bibinfo{pages}{R20} (\bibinfo{year}{1998}).

\bibitem[{\citenamefont{Poli et~al.}(2001)\citenamefont{Poli, Brecha, Roati,
  and Modugno}}]{st5}
\bibinfo{author}{\bibfnamefont{N.}~\bibnamefont{Poli}},
  \bibinfo{author}{\bibfnamefont{R.~J.} \bibnamefont{Brecha}},
  \bibinfo{author}{\bibfnamefont{G.}~\bibnamefont{Roati}}, \bibnamefont{and}
  \bibinfo{author}{\bibfnamefont{G.}~\bibnamefont{Modugno}},
  \bibinfo{journal}{Phys. Rev. A} \textbf{\bibinfo{volume}{65}},
  \bibinfo{pages}{021401(R)} (\bibinfo{year}{2001}).

\bibitem[{\citenamefont{J\'{a}uregui et~al.}(2001)\citenamefont{J\'{a}uregui,
  Poli, Roati, and Modugno}}]{st4}
\bibinfo{author}{\bibfnamefont{R.}~\bibnamefont{J\'{a}uregui}},
  \bibinfo{author}{\bibfnamefont{N.}~\bibnamefont{Poli}},
  \bibinfo{author}{\bibfnamefont{G.}~\bibnamefont{Roati}}, \bibnamefont{and}
  \bibinfo{author}{\bibfnamefont{G.}~\bibnamefont{Modugno}},
  \bibinfo{journal}{Phys. Rev. A} \textbf{\bibinfo{volume}{64}},
  \bibinfo{pages}{033403} (\bibinfo{year}{2001}).

\bibitem[{\citenamefont{Zhou et~al.}(2007)\citenamefont{Zhou, Xu, Zhou, Liu,
  and Wang}}]{st17}
\bibinfo{author}{\bibfnamefont{S.}~\bibnamefont{Zhou}},
  \bibinfo{author}{\bibfnamefont{Z.}~\bibnamefont{Xu}},
  \bibinfo{author}{\bibfnamefont{S.}~\bibnamefont{Zhou}},
  \bibinfo{author}{\bibfnamefont{L.}~\bibnamefont{Liu}}, \bibnamefont{and}
  \bibinfo{author}{\bibfnamefont{Y.}~\bibnamefont{Wang}},
  \bibinfo{journal}{Phys. Rev. A} \textbf{\bibinfo{volume}{75}},
  \bibinfo{pages}{053414} (\bibinfo{year}{2007}).

\bibitem[{\citenamefont{Paasche et~al.}(2002)\citenamefont{Paasche, Valenzuela,
  Biswas, Angelescu, and Werth}}]{st11}
\bibinfo{author}{\bibfnamefont{P.}~\bibnamefont{Paasche}},
  \bibinfo{author}{\bibfnamefont{T.}~\bibnamefont{Valenzuela}},
  \bibinfo{author}{\bibfnamefont{D.}~\bibnamefont{Biswas}},
  \bibinfo{author}{\bibfnamefont{C.}~\bibnamefont{Angelescu}},
  \bibnamefont{and} \bibinfo{author}{\bibfnamefont{G.}~\bibnamefont{Werth}},
  \bibinfo{journal}{Eur. Phys. J. D} \textbf{\bibinfo{volume}{18}},
  \bibinfo{pages}{295} (\bibinfo{year}{2002}).

\bibitem[{\citenamefont{Tommaseo et~al.}(2004)\citenamefont{Tommaseo, Paasche,
  Angelescu, and Werth}}]{st12}
\bibinfo{author}{\bibfnamefont{G.}~\bibnamefont{Tommaseo}},
  \bibinfo{author}{\bibfnamefont{P.}~\bibnamefont{Paasche}},
  \bibinfo{author}{\bibfnamefont{C.}~\bibnamefont{Angelescu}},
  \bibnamefont{and} \bibinfo{author}{\bibfnamefont{G.}~\bibnamefont{Werth}},
  \bibinfo{journal}{Eur. Phys. J. D} \textbf{\bibinfo{volume}{28}},
  \bibinfo{pages}{39} (\bibinfo{year}{2004}).

\bibitem[{\citenamefont{Higaki et~al.}(2007)\citenamefont{Higaki, Ito, Takai,
  Nakayama, Saiki, Izawa, and Okamoto}}]{st16}
\bibinfo{author}{\bibfnamefont{H.}~\bibnamefont{Higaki}},
  \bibinfo{author}{\bibfnamefont{K.}~\bibnamefont{Ito}},
  \bibinfo{author}{\bibfnamefont{R.}~\bibnamefont{Takai}},
  \bibinfo{author}{\bibfnamefont{K.}~\bibnamefont{Nakayama}},
  \bibinfo{author}{\bibfnamefont{W.}~\bibnamefont{Saiki}},
  \bibinfo{author}{\bibfnamefont{K.}~\bibnamefont{Izawa}}, \bibnamefont{and}
  \bibinfo{author}{\bibfnamefont{H.}~\bibnamefont{Okamoto}},
  \bibinfo{journal}{Hyperfine Interact} \textbf{\bibinfo{volume}{174}},
  \bibinfo{pages}{77} (\bibinfo{year}{2007}).

\bibitem[{\citenamefont{Kim et~al.}(2004)\citenamefont{Kim, Noh, Ha, and
  Jhe}}]{st14}
\bibinfo{author}{\bibfnamefont{K.}~\bibnamefont{Kim}},
  \bibinfo{author}{\bibfnamefont{H.~R.} \bibnamefont{Noh}},
  \bibinfo{author}{\bibfnamefont{H.~J.} \bibnamefont{Ha}}, \bibnamefont{and}
  \bibinfo{author}{\bibfnamefont{W.}~\bibnamefont{Jhe}},
  \bibinfo{journal}{Phys. Rev. A} \textbf{\bibinfo{volume}{69}},
  \bibinfo{pages}{033406} (\bibinfo{year}{2004}).

\bibitem[{\citenamefont{Kim et~al.}(2005)\citenamefont{Kim, Noh, and
  Jhe}}]{st15}
\bibinfo{author}{\bibfnamefont{K.}~\bibnamefont{Kim}},
  \bibinfo{author}{\bibfnamefont{H.~R.} \bibnamefont{Noh}}, \bibnamefont{and}
  \bibinfo{author}{\bibfnamefont{W.}~\bibnamefont{Jhe}},
  \bibinfo{journal}{Phys. Rev. A} \textbf{\bibinfo{volume}{71}},
  \bibinfo{pages}{033413} (\bibinfo{year}{2005}).

\bibitem[{\citenamefont{Hinds and Hughes}(1999)}]{HindsHughesReview}
\bibinfo{author}{\bibfnamefont{E.~A.} \bibnamefont{Hinds}} \bibnamefont{and}
  \bibinfo{author}{\bibfnamefont{I.~G.} \bibnamefont{Hughes}},
  \bibinfo{journal}{J. Phys. D} \textbf{\bibinfo{volume}{32}},
  \bibinfo{pages}{R119} (\bibinfo{year}{1999}).

\bibitem[{\citenamefont{{Llorente Garc\'{\i}a}
  et~al.}(2010)\citenamefont{{Llorente Garc\'{\i}a}, Darqui\'{e}, Curtis,
  Sinclair, and Hinds}}]{myVideoPaper}
\bibinfo{author}{\bibfnamefont{I.}~\bibnamefont{{Llorente Garc\'{\i}a}}},
  \bibinfo{author}{\bibfnamefont{B.}~\bibnamefont{Darqui\'{e}}},
  \bibinfo{author}{\bibfnamefont{E.~A.} \bibnamefont{Curtis}},
  \bibinfo{author}{\bibfnamefont{C.~D.~J.} \bibnamefont{Sinclair}},
  \bibnamefont{and} \bibinfo{author}{\bibfnamefont{E.~A.} \bibnamefont{Hinds}},
  \bibinfo{journal}{New J. Phys.} \textbf{\bibinfo{volume}{12}},
  \bibinfo{pages}{093017} (\bibinfo{year}{2010}).

\bibitem[{\citenamefont{{Llorente Garc\'{i}a}}(2008)}]{MyThesis}
\bibinfo{author}{\bibfnamefont{I.}~\bibnamefont{{Llorente Garc\'{i}a}}}, Ph.D.
  thesis, \bibinfo{school}{{Imperial College London}} (\bibinfo{year}{2008}).

\bibitem[{\citenamefont{Burke et~al.}(1998)\citenamefont{Burke, Bohn, Esry, and
  Greene}}]{scatteringLength}
\bibinfo{author}{\bibfnamefont{J.~P.} \bibnamefont{Burke}},
  \bibinfo{author}{\bibfnamefont{J.~L.} \bibnamefont{Bohn}},
  \bibinfo{author}{\bibfnamefont{B.~D.} \bibnamefont{Esry}}, \bibnamefont{and}
  \bibinfo{author}{\bibfnamefont{C.~H.} \bibnamefont{Greene}},
  \bibinfo{journal}{Phys. Rev. Lett.} \textbf{\bibinfo{volume}{80}},
  \bibinfo{pages}{2097} (\bibinfo{year}{1998}).

\bibitem[{\citenamefont{Guggenheim et~al.}(.)\citenamefont{Guggenheim, Mayer,
  and Tompkins}}]{PhysChem}
\bibinfo{author}{\bibfnamefont{E.~A.} \bibnamefont{Guggenheim}},
  \bibinfo{author}{\bibfnamefont{J.~E.} \bibnamefont{Mayer}}, \bibnamefont{and}
  \bibinfo{author}{\bibfnamefont{F.~C.} \bibnamefont{Tompkins}},
  \emph{\bibinfo{title}{{The International Encyclopedia of Physical Chemistry
  and Chemical Physics}}} (\bibinfo{publisher}{Pergamon Press. Oxford},
  \bibinfo{year}{.}).

\bibitem[{\citenamefont{Bird}(1994)}]{Bird94}
\bibinfo{author}{\bibfnamefont{G.~A.} \bibnamefont{Bird}},
  \emph{\bibinfo{title}{Molecular Gas Dynamics and the Direct Sim- ulation of
  Gas Flows}} (\bibinfo{publisher}{Clarendon Press, Oxford},
  \bibinfo{year}{1994}).

\bibitem[{\citenamefont{Roati et~al.}(2001)\citenamefont{Roati, Jastrzebski,
  Simoni, Modugno, and Inguscio}}]{st9}
\bibinfo{author}{\bibfnamefont{G.}~\bibnamefont{Roati}},
  \bibinfo{author}{\bibfnamefont{W.}~\bibnamefont{Jastrzebski}},
  \bibinfo{author}{\bibfnamefont{A.}~\bibnamefont{Simoni}},
  \bibinfo{author}{\bibfnamefont{G.}~\bibnamefont{Modugno}}, \bibnamefont{and}
  \bibinfo{author}{\bibfnamefont{M.}~\bibnamefont{Inguscio}},
  \bibinfo{journal}{Phys. Rev. A} \textbf{\bibinfo{volume}{63}},
  \bibinfo{pages}{052709} (\bibinfo{year}{2001}).

\bibitem[{\citenamefont{Barrett}(2002)}]{st19}
\bibinfo{author}{\bibfnamefont{M.~D.} \bibnamefont{Barrett}}, Ph.D. thesis,
  \bibinfo{school}{Georgia Institute of Technology} (\bibinfo{year}{2002}).

\end{thebibliography}

\end{document}